\newcommand{\cl}{\centerline}
\documentstyle[12pt]{article}
\begin{document}
\hfill{CCUTH-95-02}\par
\setlength{\textwidth}{5.2in}
\setlength{\textheight}{7.5in}
\setlength{\parskip}{0.0in}
\setlength{\baselineskip}{18.2pt}
\vfill
\cl{\large{{\bf Resummation of Large Corrections}}}\par
\cl{\large{{\bf in Inclusive $B$ Meson Decays}}}\par
\vskip 0.5cm
\cl{Hsiang-nan Li}
\vskip 0.5cm
\cl{Department of Physics, National Chung-Cheng University,}
\cl{Chia-Yi, Taiwan, R.O.C.}
\vskip 0.5cm
\cl{\today }
\vskip 2.0cm
%\baselineskip=2.0\baselineskip

\cl{\bf Abstract}

We show that the resummation technique developed recently for heavy-light
systems and the conventional approach based on the factorization of Wilson
loops are equivalent. The two methods lead to the same results of the
resummation of large corrections in inclusive $B$ meson decays.

\vfill
\newpage

\cl{\large \bf 1. Introduction}
\vskip 0.5cm

Perturbative QCD (PQCD) including Sudakov effects \cite{BS,LS} has been
proposed to be an appropriate approach to exclusive heavy meson decays in
the maximal recoil region of final-state hadrons \cite{YS,LY,L1,DJK}. The
progress is attributed to the resummation of large radiative corrections in
heavy-light systems such as a $B$ meson containing a light valence quark,
through which important higher-order and higher-power contributions are
included into factorization theorems. Recently, the resummation technique
was extended to the inclusive semileptonic decay $B\to X_u\ell\nu$
\cite{LY2}, whose behavior near the high end of charged lepton energy is
crucial to the extraction of the Cabibbo-Kobayashi-Maskawa matrix elements
$|V_{ub}|$ in the standard model and to the detection of new physics. It
has been observed that large perturbative corrections appear at the end
point of the spectrum \cite{BSUV,ACC,FJM}, which must be organized to all
orders in order to have a reliable PQCD analysis.

The above approach to the resummation of large corrections in inclusive
processes differs from the conventional one. In the study of deep inelastic
scatterings and Drell-Yan productions \cite{CT}, large soft corrections
are factorized systematically into path-ordered exponentials, or Wilson
lines \cite{KM,CC,FST} along the classical trajectories of quarks.
Korchemsky and Sterman \cite{KS} applied this Wilson-loop formalism to the
study of end-point singularities in inclusive $B$ meson decays. The Wilson
lines are absorbed into a heavy quark distribution function $S$, and give
its evolution governed by renormalization group (RG) equations. The emitted
light quark produces a jet of collinear particles at the end point due to
its large energy and small invariant mass \cite{KS}.

In this letter we shall show that the resummation technique is equivalent
to the Wilson-loop formalism, if relevant evolution scales are chosen
properly. The two methods give the same resummation results for the
inclusive decays $B\to X_s\gamma$ and $B\to X_u\ell\nu$ up to next-to-leading
logarithms. Therefore, this work provides a non-trivial check on the two
different approaches.

\vskip 2.0cm
\centerline{\large \bf 2. The Radiative Decay $B\to X_s\gamma$}
\vskip 0.5cm

We apply the resummation technique to the inclusive radiative decay
$B\to X_s\gamma$, which occurs through the transition $b\to s\gamma$
described by an effective Hamiltonian \cite{SVZ}. The basic factorization of
this process is shown in fig.~1a, where the bulb represents the $b$ quark
distribution function $S(k)$, $k$ being the momentum carried by light
partons in the $B$ meson, $k^2\approx 0$. $k$ has a plus component $k^+$ and
small transverse components ${\bf k}_T$, which serve as the infrared cutoff
of loop integrals for radiative corrections. The momentum of the $b$ quark
is then $P-k$, $P=m_B/\sqrt{2}(1,1,{\bf 0})$ being the $B$ meson momentum
in terms of light-cone components, which satisfies the on-shell condition
$(P-k)^2\approx m_b^2$. Here $m_B$ and $m_b$ are the $B$ meson mass and the
$b$ quark mass, respectively. The $b$ quark decays into a real photon of
momentum $q$ and a $s$ quark of momentum $P_s$, which is regarded as being
light. Assuming that the nonvanishing component of $q$ is $q^+$, we have
$P_s=P-k-q=(P^+-k^+-q^+,P^-,-{\bf k}_T)$.

We work in axial gauge $n\cdot A=0$ for the analysis below, $n$ being a
gauge vector. In the end-point region with $q^+\to P^+$, the $s$ quark has
a large minus component $P^-$ but a small invariant mass $P_s^2$. Hence,
photon vertex corrections and self-energy corrections to the $s$ quark in
fig.~1b give double logarithms from the overlap of collinear and soft
divergences. In the leading regions with loop momentum $l$ parallel to $P_s$
and with soft $l$, the integrand associated with the photon vertex correction
has the partial expression
\begin{eqnarray}
\frac{(\not P-\not k+\not l+m_b)\gamma_\lambda(\not P+m_B)}
{[(P-k+l)^2-m_b^2]}\approx
\frac{2P_\lambda-\gamma_\lambda(\not P-m_b)+\not l\gamma_\lambda}
{2P\cdot l}(\not P+m_B)\;,
\label{4be}
\end{eqnarray}
where the factor $(\not P+m_B)$ is the matrix structure of $S$, and those
terms proportional to $k$ have been neglected. The second term in the
numerator of eq.~(\ref{4be}) is suppressed by power $(m_B-m_b)/m_B$. The
third term is negligible for soft $l$. For $l//P_s$, it can be easily shown
that the trace of $\gamma$-matrices associated with fig.~1a, which contains
the third term, vanishes.

Therefore, in the leading regions of $l$ radiated gluons are detached from
the $b$ quark and collected by an eikonal line described by the Feynman rule
$P_\lambda /P\cdot l$. The factor $1/P\cdot l$ is associated with the
eikonal propagator, and the numerator $P_\lambda$ is assigned to the vertex,
where a gluon attaches the eikonal line. This observation is consistent with
the flavor symmetry in the heavy quark effective theory \cite{G}. We then
absorb this type of radiative corrections into a jet function $J(P_s)$,
which is now factorized out of the process. This absorption is reasonable
because the soft divergences in fig.~1b cancel asymptotically as explained
later, and the double logarithmic corrections are mainly collinear.

Self-energy corrections to the $b$ quark and loop corrections with real
gluons connecting the two $b$ quarks in fig.~1c contain only single soft
logarithms, and are grouped into the distribution function, which is
supposed to be dominated by soft dynamics. Remaining important corrections
from hard gluons are absorbed into a hard scattering amplitude $H$. At last,
the factorization formula for the spectrum of the decay $B\to X_s\gamma$ is
written as
\begin{equation}
\frac{d \Gamma_\gamma}{d E_\gamma}=
\int d k^+d^2 {\bf k}_TS(k,\mu)H(P-k,\mu)J(P_s,\mu)\;,
\label{fac}
\end{equation}
which is described graphically in fig.~2. Here $E_\gamma=q^+/\sqrt{2}$ is
the photon energy, and $\mu$ is the factorization and renormalization scale.

The Wilson-loop formalism for the decay $B\to X_s\gamma$ \cite{KS} is
summarized in fig.~3. Soft gluons radiated from the $b$ quark are collected
by the eikonal lines along the classical trajectories of the $b$ and $s$
quarks, which form a Wilson loop, and are absorbed into $S$. Since the $s$
quark is almost on-shell in the end-point region, the Wilson loop introduces
extra collinear singularities into the distribution function, which are not
consistent with the dominant soft dynamics in the $B$ meson. Note that the
light $s$ quark is not replaced by an eikonal line in our treatment. Hence,
both $S$ and $J$ in \cite{KS} contain double logarithms, which are
treated by RG methods. However, due to the presence of double logarithms,
relevant anomalous dimensions involved in RG equations must be accompanied
by large logarithms \cite{KM}, such that perturbative calculation is not
reliable.

For the purpose of demonstrating the equivalence of our approach to the
Wilson-loop one, it is convenient to choose a time-like gauge vector
$n=P$, instead of a space-like $n$ as in \cite{LY,L1,LY2}, under which
radiated gluons do not attach the eikonal lines associated with the $b$
quarks. This choice of $n$ does not bring new pinch singularities into $S$.

We drop the intrinsic ${\bf k}_T$ dependence in $S$, and Fourier transform
eq.~(\ref{fac}) into $b$ space, $b$ being a conjugate variable to $k_T$,
\begin{equation}
\frac{d \Gamma_\gamma}{d E_\gamma}=
\int d k^+\frac{d^2 {\bf b}}{(2\pi)^2}S(k^+,\mu){\tilde H}
(P^+-k^+,{\bf b},\mu){\tilde J}(P_s^+,P_s^-,{\bf b},\mu)\;.
\label{fac1}
\end{equation}
It has been shown that integrands for soft corrections are proportional to
$1-e^{i{\bf l}_T\cdot{\bf b}}$ \cite{BS,LY2}, which vanish in the asymptotic
region with $b\to 0$ as stated before.

$\tilde J$ depends on $P_s$ and $n$ through the ratio $(P_s\cdot n)^2/n^2$
due to the scale invariance in $n$ of the gluon propagator in axial gauge,
\begin{equation}
N^{\mu\nu}(l)=\frac{-i}{l^2}\left(g^{\mu\nu}-\frac{n^\mu l^\nu+n^\nu l^\mu}
{n\cdot l}+n^2\frac{l^\mu l^\nu}{(n\cdot l)^2}\right)\;.
\end{equation}
To sum up the double logarithms in $\tilde J$, we consider the derivative
$d\tilde J/d\ln P_s^-$ \cite{LY,LY2}, $P_s^-$ being regarded as a variable
from now on. Replacing $d/d \ln P_s^-$ by $d/d n$ using a chain rule, we
have
\begin{equation}
\frac{d}{d \ln P_s^-}{\tilde J}=-\frac{n^2}{v\cdot n}v_{\alpha}
\frac{d}{d n_{\alpha}}{\tilde J}
\label{qn}
\end{equation}
with the vector $v=(0,1,{\bf 0})$. $d/d n_{\alpha}$ operates only on a gluon
propagator, and gives \cite{BS}
\begin{equation}
\frac{d}{d n_{\alpha}}N^{\mu\nu}=-\frac{1}{l\cdot n}
(N^{\mu\alpha}l^{\nu}+N^{\nu\alpha}l^{\mu})\;.
\label{dn}
\end{equation}
Combining eqs.~(\ref{qn}) and (\ref{dn}), we find that a gluon vertex,
after contracted with $l$ that locates at both ends of the differentiated
gluon line, has been replaced by a new vertex, described by the Feynman rule
$\frac{n^2}{v\cdot n l\cdot n}v_{\alpha}$. This new vertex will be
represented by a square. Adding together all diagrams with different
differentiated gluon lines and using the Ward identity \cite{BS,CS}, this
square moves to the outermost ends of $\tilde J$.

The gluon momentum $l$ flowing through the square vertex does not lead to
collinear divergences because of the denominator $n\cdot l$. The leading
regions of $l$ are then soft and ultraviolet, in which the subdiagram
containing the square can be factorized. We then derive a differential
equation,
\begin{equation}
\frac{d}{d \ln P_s^-}{\tilde J}=
2\left[{\cal K}(\alpha_s(\mu))+{\cal G}(\alpha_s(\mu))\right]{\tilde J}\;,
\label{dfe}
\end{equation}
as shown in fig.~4a, where the eikonal lines disappear due to the choice
$n=P$. The factor 2 counts the two ends of $\tilde J$. The
functions ${\cal K}$ and ${\cal G}$ collect the soft and ultraviolet
divergences in the subdiagram, respectively. Lowest-order diagrams of
$\cal K$ are exhibited in fig.~4b, and those of $\cal G$ are in fig.~4c.

A straightforward calculation gives ${\cal K}={\rm fig.\;4b}-\delta {\cal K}$
and ${\cal G}={\rm fig.\;4c}-\delta {\cal G}$, $\delta {\cal K}$ and
$\delta {\cal G}=-\delta {\cal K}$ being the corresponding counterterms.
Since $\cal K$ contains only single soft logarithms and $\cal G$ contains
only single ultraviolet logarithms, they can be treated by RG methods:
\begin{equation}
\mu\frac{d}{d\mu}{\cal K}=-\lambda_{\cal K}=
-\mu\frac{d}{d\mu}{\cal G}\;.
\label{kg}
\end{equation}
$\lambda_{\cal K}=\mu\frac{d\delta {\cal K}}{d\mu}$ is the anomalous
dimension of ${\cal K}$, whose expression up to two loops is given by
\cite{BS,LS}
\begin{equation}
\lambda_{\cal K}=\frac{\alpha_s}{\pi}{\cal C}_F+\left(\frac{\alpha_s}{\pi}
\right)^2{\cal C}_F\left[{\cal C}_A\left(\frac{67}{36}
-\frac{\pi^{2}}{12}\right)-\frac{5}{18}n_{f}\right]
\label{lk}
\end{equation}
with $n_{f}$ the number of quark flavors, and ${\cal C}_F=4/3$ and
${\cal C}_A=3$ the color factors. $\lambda_{\cal K}$ is independent of the
gauge vector $n$, and thus eq.~(\ref{lk}) is identical to that derived from
a space-like $n$ in \cite{LY2}.

There are two $P_s^-$-dependent scales involved in $\tilde J$: one is
$P_s^-$ itself, which serves as the lower limit of $\mu$, and the other
comes from the invariant $P\cdot P_s\approx P^+P_s^-$. Solving
eq.~(\ref{kg}), we have
\begin{eqnarray}
{\cal K}+{\cal G}&=&
{\cal K}(\alpha_s(P_s^-))+{\cal G}\left(\alpha_s\left(\sqrt{P^+P_s^-}\right)
\right)+\int_{P_s^-}^{\sqrt{P^+P_s^-}}\frac{d{\mu}}{\mu}
\lambda_{\cal K}(\alpha_s({\mu}))\;,
\nonumber \\
&\approx &
\int_{P_s^-}^{\sqrt{P^+P_s^-}}\frac{d{\mu}}{\mu}
\lambda_{\cal K}(\alpha_s({\mu}))\;.
\label{skg}
\end{eqnarray}
In the second expression we have neglected the initial condition of
${\cal K}+{\cal G}$. Substituting eq.~(\ref{skg}) into (\ref{dfe}), we
derive
\begin{eqnarray}
{\tilde J}(P_s^+,P_s^-,{\bf b},\mu)&=&
\exp\left[-2\int_{1/b}^{P^-}\frac{d P_s^-}{P_s^-}
\int_{P_s^-}^{\sqrt{P^+P_s^-}}\frac{d{\bar \mu}}{\bar \mu}
\lambda_{\cal K}(\alpha_s({\bar \mu}))\right]
\nonumber \\
& &\times {\tilde J}(P_s^+,{\bf b},\mu)\;,
\label{sj}
\end{eqnarray}
where we have taken the infrared cutoff $1/b$ (conjugate to $k_T$) as the
lower bound of $P_s^-$ and the largest scale $P^-$ involved in the process
as the upper bound. The four scales appearing in eq.~(\ref{sj}) satisfy
the ordering $P^->\sqrt{P^+P_s^-}>P_s^->1/b>\Lambda_{\rm QCD}$, which
justifies the neglect of the initial condition of ${\cal K}+{\cal G}$.

The function ${\tilde J}(P_s^+,{\bf b},\mu)$ obeys the RG equation,
\begin{equation}
{\cal D}{\tilde J}(P_s^+,{\bf b},\mu)=-2\lambda_J {\tilde J}(P_s^+,
{\bf b},\mu)\;,
\label{uj}
\end{equation}
with ${\cal D}=\mu\frac{\partial}{\partial \mu}+\beta(g)\frac{\partial}
{\partial g}$ and $\lambda_J=-\alpha_s/\pi$ the quark anomalous dimension
in axial gauge. The invariant mass of the jet, $P_s^2\approx 2P_s^+P^-
\approx 2P^-/b$, if assuming that the small $P_s^+$ is of the same order as
$1/b$, is a natural initial scale of the evolution. The solution to
eq.~(\ref{uj}) is then given by
\begin{equation}
{\tilde J}(P_s^+,{\bf b},\mu)=
\exp\left[-\int_{P^-/b}^{\mu^2}\frac{d {\bar \mu}^2}
{{\bar \mu}^2}\lambda_J (\alpha_s({\bar \mu})) \right]{\tilde J}
(P_s^+,{\bf b},\sqrt{P^-/b})\;.
\label{ujj}
\end{equation}

To sum up next-to-leading logarithms, we have to organize the single soft
logarithms in $S$. The RG equation for $S$ is written as
\begin{equation}
{\cal D}S(k^+,\mu)=-\lambda_S S(k^+,\mu)\;,
\label{us}
\end{equation}
where $\lambda_S=-(\alpha_s/\pi){\cal C}_F$ is the anomalous dimension
derived from fig.~1c with the eikonal approximation. We allow $S$ to evolve
to the infrared cutoff $1/b$, and obtain
\begin{equation}
S(k^+,\mu)=\exp\left[-\int_{1/b}^{\mu}\frac{d {\bar \mu}}
{\bar \mu}\lambda_S (\alpha_s({\bar \mu})) \right]S(k^+,1/b)\;.
\label{uss}
\end{equation}
The initial condition $S(k^+,1/b)$ has a nonperturbative origin arising
from soft QCD dynamics in the $B$ meson.

Since the differential decay rate is independent of $\mu$, the RG equation
of the hard scattering $\tilde H$ is written as
\begin{equation}
{\cal D}{\tilde H}(\mu)=(2\lambda_J+\lambda_S) {\tilde H}(\mu)\;,
\label{sh}
\end{equation}
which is solved to give
\begin{equation}
{\tilde H}(\mu)=\exp\left[-\int_{\mu^2}^{(P^-)^2}\frac{d {\bar \mu}^2}
{{\bar \mu}^2}\lambda_J(\alpha_s({\bar \mu}))
-\int_{\mu}^{P^-}\frac{d {\bar \mu}}
{\bar \mu}\lambda_S (\alpha_s({\bar \mu}))
\right]{\tilde H}(P^-)\;.
\label{uh}
\end{equation}
We have chosen the largest scale $P^-$ as the upper bound of the integral.

Combining eqs.~(\ref{sj})-(\ref{uh}), we derive the differential decay rate
\begin{eqnarray}
\frac{d \Gamma_\gamma}{d E_\gamma}&=&
\int d k^+\frac{d^2 {\bf b}}{(2\pi)^2}S(k^+,1/b){\tilde H}
(P^+-k^+,{\bf b},P^-){\tilde J}(P_s^+,{\bf b},\sqrt{P^-/b})
\nonumber \\
& &\times\exp(-s_\gamma(P^-,1/b))\;,
\label{fae}
\end{eqnarray}
with the Sudakov exponent
\begin{eqnarray}
s_\gamma&=&2\int_{1/b}^{P^-}\frac{d P_s^-}{P_s^-}
\int_{P_s^-}^{\sqrt{P^+P_s^-}}\frac{d{\bar \mu}}{\bar \mu}
\lambda_{\cal K}(\alpha_s({\bar \mu}))
\nonumber \\
& &+\int_{1/b}^{P^-}\frac{d {\bar \mu}}
{\bar \mu}\lambda_S (\alpha_s({\bar \mu}))
+\int_{P^-/b}^{(P^-)^2}\frac{d {\bar \mu}^2}
{{\bar \mu}^2}\lambda_J(\alpha_s({\bar \mu})) \;.
\label{fs}
\end{eqnarray}
Employing the variable changes $P_s^-=yP^-$ in the first term,
${\bar \mu}=yP^-$ in the second term and ${\bar \mu}^2=yP^{-2}$ in the
last term, eq.~(\ref{fs}) reduces to
\begin{equation}
s_\gamma=\int_{1/bP^-}^1\frac{d y}{y}\left[
2\int_{yP^-}^{\sqrt{y}P^-}\frac{d{\bar \mu}}{\bar \mu}
\lambda_{\cal K}(\alpha_s({\bar \mu}))
+\lambda_S(\alpha_s(yP^-))+\lambda_J(\alpha_s(\sqrt{y}P^-))\right]\;.
\label{fs1}
\end{equation}
The corresponding expression for the exponent in \cite{KS} is quoted below:
\begin{equation}
s_N=\int_{n_0/N}^{1}\frac{d y}{y}\left[
2\int_{yM}^{\sqrt{y}M}\frac{d k_t}{k_t}
\Gamma_{\rm cusp}(\alpha_s(k_t))+\Gamma(\alpha_s(yM))
+\gamma(\alpha_s(\sqrt{y}M))\right]\;,
\label{fks}
\end{equation}
where $s_N$ describes the evolution of the $N$-th moment of the differential
decay rate, with $M$ the $b$ quark mass, the constant $n_0=e^{-\gamma_E}$,
$\gamma_E$ being the Euler constant, and
$\Gamma_{\rm cusp}=\lambda_{\cal K}$, $\Gamma=-(\alpha_s/\pi){\cal C}_F$ and
$\gamma=-\alpha_s/\pi$ the anomalous dimensions. The two-loop expressions
for $\Gamma_{\rm cusp}$ and $\Gamma$ can be found in \cite{KM,KR1}.

Comparing eq.~(\ref{fs1}) to (\ref{fks}), we easily identify the equality
of the anomalous dimensions:
\begin{eqnarray}
\lambda_{\cal K}=\Gamma_{\rm cusp}\;,\;\;\;\;
\lambda_S=\Gamma\;,\;\;\;\;\lambda_J=\gamma\;.
\label{re}
\end{eqnarray}
The two expressions $s_\gamma$ and $s_N$ are basically the same except the
lower bounds of $y$. The scale $1/b$ comes from the inclusion of transverse
degrees of freedom in our formalism, and the $N$ dependence is due to the
moment analysis in \cite{KS}.

\vskip 2.0cm

\centerline{\large \bf 3. The Semileptonic Decay $B\to X_u\ell\nu$}
\vskip 0.5cm
The analysis of the inclusive semileptonic decay $B\to X_u\ell\nu$
is similar to that of the radiative decay $B\to X_s\gamma$. Large double
logarithmic corrections are also observed near the high end of the charged
lepton spectrum. We choose the $B$ meson momentum $P$ and the light parton
momentum $k$ as in Section 2. The lepton momentum $P_\ell$ has only a large
plus component $P_\ell^+$. The neutrino carries the momentum $P_\nu$,
$P_\nu^2=0$. The $u$ quark, emitted by the $b$ quark through the transition
${\bar u}\gamma^\mu b$, has the momentum $P_u=P-k-q$ with the lepton pair
momentum $q=P_\ell+P_\nu =(P_\ell^++P_\nu^+,P_\nu^-,{\bf P}_{\nu T})$. When
$P_\ell^+$ reaches its maximum $P^+$, implying $P_\nu^+\approx 0$ and
${\bf P}_{\nu T}\approx 0$, the $u$ quark has a small invariant mass, and
moves fast in the minus direction with $P_u^-=P^--P_\nu^-$. The situation
is then similar to the end-point region of the radiative decay
$B\to X_s\gamma$.

Repeating the resummation procedures in Section 2, we derive the
corresponding Sudakov exponent $s_\ell$, which is similar to $s_\gamma$ but
with the largest scale $P^-$ replaced by $(1-x_\nu)P^-$, $x_\nu=P_\nu^-/P^-$.
$s_\ell$ is given by
\begin{eqnarray}
s_\ell&=&2\int_{1/b}^{(1-x_\nu)P^-}\frac{d P_u^-}{P_u^-}
\int_{P_u^-}^{\sqrt{P^+P_u^-}}\frac{d{\bar \mu}}{\bar \mu}
\lambda_{\cal K}(\alpha_s({\bar \mu}))
\nonumber \\
& &+\int_{1/b}^{(1-x_\nu)P^-}\frac{d {\bar \mu}}
{\bar \mu}\lambda_S (\alpha_s({\bar \mu}))
+\int_{(1-x_\nu)P^-/b}^{(1-x_\nu)^2(P^-)^2}\frac{d {\bar \mu}^2}
{{\bar \mu}^2}\lambda_J(\alpha_s({\bar \mu})) \;,
\nonumber \\
&=&\int_{1/bP^-}^{1-x_\nu}\frac{d y}{y}\left[
2\int_{yP^-}^{\sqrt{y}P^-}\frac{d{\bar \mu}}{\bar \mu}
\lambda_{\cal K}(\alpha_s({\bar \mu}))+\lambda_S(\alpha_s(yP^-))\right]
\nonumber \\
& &+\int_{(1-x_\nu)/bP^-}^{(1-x_\nu)^2}\frac{dy}{y}
\lambda_J(\alpha_s(\sqrt{y}P^-))\;.
\label{fss}
\end{eqnarray}
Note the change in the $y$-integration limits compared to eq.~(\ref{fs1}).
\vskip 2.0cm

\centerline{\large \bf 4. Conclusion}
\vskip 0.5cm

In this letter we have shown in detail that the resummation technique
is indeed equivalent to the conventional Wilson-loop formalism. The two
approaches lead to the same resummation results in the end-point region of
inclusive $B$ meson decays, except that the large logarithm appears as
$\ln bP^-$ in our expression and as $\ln N$ in \cite{KS}. This equivalence
provides a justification of our PQCD analysis of $B$ meson decays including
Sudakov effects in previous works \cite{LY,L1,LY2}.

We have neglected the initial condition of ${\cal K}+{\cal G}$ in
demonstrating the equivalence, which in fact contributes to
next-to-leading logarithms. However, we believe that physical predictions
should be insensitive to the distinction in next-to-leading logarithms. It
is known that a substitution of the scale $P^-$ by another one $CP^-$,
with $C$ a constant of order 1, is allowed in resummation \cite{BS,CS}.
This constant $C$ in the integral involving $\lambda_{\cal K}$ then makes a
difference at the level of next-to-leading logarithms. Hence, if the
initial condition of ${\cal K}+{\cal G}$ is included, predictions will be
modified only slightly. Certainly, this issue needs more careful study.
Other interesting subjects, such as the gauge dependence of resummation
and the application of our technique to deep inelastic scatterings and
Drell-Yan productions, will be discussed in a seperate work \cite{L3}.

\vskip 0.5cm
This work was supported by the National Science Council of R.O.C. under
Grant No. NSC84-2112-M194-006.

\newpage
\cl{\large \bf Figure Captions}
\vskip 0.5cm

\noindent
{\bf Fig. 1.} (a) Basic factorization for the decay
$B\to X_s\gamma$, and (b) and (c) $O(\alpha_s)$ radiative corrections.
\vskip 0.5cm

\noindent
{\bf Fig. 2.} Factorization for the decay $B\to X_s\gamma$ in our formalism
for a general gauge vector $n$.
\vskip 0.5cm

\noindent
{\bf Fig. 3.} Factorization for the decay $B\to X_s\gamma$ in the
Wilson-loop formalism.
\vskip 0.5cm

\noindent
{\bf Fig. 4.} (a) Graphic representation of eq.~(\ref{dfe}) for a gauge
vector $n=P$, and ${\cal O}(\alpha_s)$ diagrams for (b) the function
${\cal K}J$ and for (c) the function ${\cal G}J$.
\vskip 0.5cm


\vskip 2.0cm

\newpage
\begin{thebibliography}{99}
\bibitem{BS} J. Botts and G. Sterman, Nucl. Phys. B325 (1989) 62.
\bibitem{LS} H.-n. Li and G. Sterman, Nucl. Phys. B381 (1992) 129;
H.-n. Li, Phys. Rev. D48 (1993) 4243.
\bibitem{YS} R. Akhoury, G. Sterman and Y.-P. Yao, Phys. Rev. D50 (1994)
358.
\bibitem{LY} H.-n. Li and H.L. Yu, Phys. Rev. Lett. 74 (1995) 4388;
Phys. Lett. B353 (1995) 301; H.-n. Li, Phys. Lett. B348 (1995) 597.
\bibitem{L1} H.-n. Li, Phys. Rev. D52 (1995) 3958.
\bibitem{DJK} M. Dahm, R. Jakob and P. Kroll, preprint WU-B-95-06.
\bibitem{LY2} H.-n. Li and H.L. Yu, preprint CCUTH-95-04.
\bibitem{BSUV} I.I. Bigi, M.A. Shifman, N.G. Uraltsev and A.I. Vainshtein,
Int. J. Mod. Phys. A9 (1994) 2467.
\bibitem{ACC} G. Altarelli, N. Cabibbo, G. Corbo, L. Maiani and G.
Martinelli, Nucl. Phys. B208 (1982) 365.
\bibitem{FJM} A.F. Falk, E. Jenkins, A.V. Manohar and M.B. Wise, Phys. Rev.
D49 (1994) 4553.
\bibitem{CT} S. Catani and L. Trentadue, Nucl. Phys. B327 (1989) 323;
B353 (1991) 183.
\bibitem{KM} G.P. Korchemsky and G. Marchesini, Nucl. Phys. B406 (1993)
225; G.P. Korchemsky, Mod. Phys. Lett. A4 (1989) 1257.
\bibitem{CC} S. Catani and M. Ciafaloni, Nucl. Phys. B236 (1984) 61; B249
(1985) 301; S. Catani and M. Ciafaloni and G. Marchesini, Nucl. Phys. B264
(1986) 558.
\bibitem{FST} J. Frenkel, J.G.M. Gatherall and J.C. Taylor, Nucl. Phys.
B233 (1984) 307; J. Frenkel, P. Sorensen and J.C. Taylor, Z. Phys. C35
(1987) 361.
\bibitem{KS} G.P. Korchemsky and G. Sterman, Phys. Lett. B340 (1994) 96.
\bibitem{SVZ} M.A. Shifman, A.I. Vainshtein and V.I. Zakharov, Phys. Rev.
D18 (1978) 2583.
\bibitem{G} H. Georgi, Phys. Lett. B240 (1990) 447.
\bibitem{CS} J.C. Collin and D.E. Soper, Nucl. Phys. B193 (1981) 381.
\bibitem{KR1} G.P. Korchemsky and A.V. Radyushkin,
Nucl. Phys. B283 (1987) 342.
\bibitem{L3} H.-n. Li, in preparation.
\end{thebibliography}
\end{document}